\documentclass[journal=jcp,manuscript=article,twosides,a4paper]{achemso}
\usepackage[version=3]{mhchem} % Formula subscripts using \ce{}
\usepackage[english]{babel}
\usepackage[utf8]{inputenc}
\usepackage[T1]{fontenc}
\usepackage{color}
\usepackage{rotating}
\usepackage{tikz}
\usetikzlibrary{shapes,arrows}
\usetikzlibrary{arrows.meta}
\usepackage{amssymb}
\usepackage[font=small,labelfont=bf,tableposition=top]{caption}
\usepackage{enumerate}
\usepackage{array,dcolumn}
\usepackage{xspace}
\usepackage{graphicx}
\usepackage{lscape}
\usepackage{rotating}
\usepackage{graphics}
\usepackage{amsmath}
\usepackage{amsfonts}
\usepackage{eso-pic}
\usepackage{epstopdf}
\usepackage{multimedia}
\usepackage{braket}
\usepackage{color}
\usepackage{times} 
\usepackage{bm}
\usepackage{txfonts} 
\usepackage{subcaption}
\usepackage{soul}
\usepackage{nicefrac} % For comparison

\newcommand{\lcpq}{Laboratoire de Chimie et Physique Quantiques, CNRS, \\
Universit\'e de Toulouse 3 Paul Sabatier, \\
118 route de Narbonne, 31062 France}
\newcommand{\etsf}{European Theoretical Spectroscopy Facility (ETSF)}
\newcommand{\unibo}{Universit\`a di Bologna, \\
Via Irnerio 33, 40126 Bologna, Italy}
\newcommand{\ehu}{Polimero eta Material Aurreratuak: Fisika, Kimika eta Teknologia saila, Kimika Fakultatea, Euskal Herriko Unibertsitatea, UPV/EHU,
and Donostia International Physics Center (DIPC). 
P.K. 1072, 20080 Donostia, Euskadi, Spain}

\title{The Wigner localization of interacting electrons in a one-dimensional harmonic potential} 

\author{Xabier Telleria-Allika}
\affiliation{\ehu}
\author{Miguel Escobar Azor}
\affiliation{\lcpq}
\author{Gr\'egoire Fran\c{c}ois}
\affiliation{\lcpq}
\author{Gian Luigi Bendazzoli}
\affiliation{\unibo}
\author{Jon M. Matxain}
\affiliation{\ehu}
\author{Xabier Lopez}
\affiliation{\ehu}
\email{xabier.lopez@ehu.eus}
\author{Stefano Evangelisti}
\email{stefano.evangelisti@irsamc.ups-tlse.fr}
\affiliation{\lcpq}
\author{J. Arjan Berger}
\email{arjan.berger@irsamc.ups-tlse.fr}
\affiliation{\lcpq}
\affiliation{\etsf}
\date{\today}
\setkeys{acs}{articletitle = true}
\begin{document}

\begin{abstract}

In this work we study the Wigner localization of interacting electrons that are confined to a quasi-one-dimensional harmonic potential using accurate quantum chemistry approaches. We demonstrate that the Wigner regime can be reached using small values of the confinement parameter. To obtain physical insight in our results we analyze them with a semi-analytical model for two electrons.
Thanks to electronic-structure properties such as the one-body density and the particle-hole entropy, we are able to define a path that connects the Wigner regime to the Fermi-gas regime by varying the confinement parameter. 
In particular, we show that the particle-hole entropy as a function of the confinement parameter smoothly connects the two regimes. Moreover, it exhibits a maximum that could be interpreted as the transition point between the localized and delocalized regimes.
%whose position depends on the number of electrons inside the harmonic potential.
\end{abstract}

\section{Introduction}

By means of basic concepts concerning quantum mechanics governing interactions of systems composed of electrons, in 1934 Wigner predicted that at very low density an electron gas crystallizes, i.e., electrons localize on crystal lattice sites~\cite{wigner_paper}. In fact, since the Coulomb repulsion per electron scales as $1/r_s$ (where the Wigner-Seitz radius $r_s$ is the average distance between electrons) and the average kinetic energy per electron scales as $1/r_s^2$, it can be understood that at low density (large $r_S$), the Coulomb repulsion dominates the kinetic energy. Therefore, at sufficient low densities, electrons can minimize the total energy of the system by getting localized; this phase is called a Wigner crystal. Instead, at high densities (small $r_S$) the kinetic energy will dominate the Coulomb repulsion and the electrons will form a Fermi gas.
Transitions happening between these two quantum states of matter have experimentally been observed in several recent works \cite{he_wigner_2009,metling_criterion,quantum_melting,quantum_melting_2,imagin_wigner,quantum_phase_diag,wigner_molecule_2019} and, interestingly, a substrate for Wigner crystals seems to be bilayer materials in which low densities are obtained by means of emerging Moir\'e patterns \cite{bilayer_1995,bilayer_graphene_insu,nature_2021,gen_wigner_crys}.

Localized electronic states can emerge also in the context of few-electrons systems, when the confinement of the electrons is rather weak~\cite{wigner_mol_1999,wigner_mol_2007,wigner_mol_2009, 1D_Wigner_Alejandro}. 
Because the underlying physics is similar to that of Wigner crystals, these systems are often referred to as Wigner molecules.
These systems display a rich variety of electronic states, consisting of several ground-state spin multiplicities and multi-determinantal features~\cite{varga:01prb,cioslowski:12jcp,cioslowski:14jcp}. 
Wigner molecules have also been observed experimentally \cite{zig_zag_1_d_2018,1d_few_el_crys_exp_2019}.
Since strong electron correlation drives the Wigner localization, the \emph{ab initio} study of this phenomenon requires accurate quantum-chemistry approaches in order to obtain highly accurate energies and wave functions in both the low-correlation and the high-correlation regimes~\cite{sahni_2014,gori-giorgi_savin_2009,zhu_trickey_2006,PhysRevB.84.115302}. 
We notice that, besides being of potential technological interest, Wigner molecules are also studied to understand many-body interactions. 
In particular, they have been repeatedly used in the calibration of electronic structure methods, since they provide insight into both the dynamical and non-dynamical electron correlation regimes~\cite{ramos-cordoba:16pccp,ramos-cordoba:17jctc,via-nadal:19jpcl} that pose a great challenge for current computational methods.~\cite{hessler:82prl,laufer:86pra,kais:93jcp,filippi:94jcp,huang:97pra,taut:98jpb,qian:98pra,ivanov:99jcp,ludena:00bc,zhu:06jcp,cioslowski:11jctc,cioslowski:15jcp,rodriguez:17pccp,rodriguez:17pccp2}
Such calibration has been made possible because of the recent availability of highly accurate analytical data and benchmark results.~\cite{cioslowski:00jcp,varga:01prb,cioslowski:11jcp,amovili:11pra,cioslowski:12jcp,cioslowski:13jcp,strasburger:16jcp,cioslowski:17jcp,cioslowski:18jcp}
Some of us have recently studied the electron localization in Wigner molecules using the exact diagonalisation of the Hamiltonian and configuration interaction techniques.~\cite{1D_Wigner_Alejandro, wigner_mol_ring_2019,2_3_d_wigner_crystal} In those works the electrons were either implicitly confined by the domain of the basis functions or explicitly by confining the electrons to a circle or a Clifford torus.~\cite{Tavernier_2020,Tavernier_2021}

In the present work, we confine the electrons with a one-dimensional harmonic potential and we will study the electronic properties of the system, and in particular the Wigner localization, when varying the harmonic constant. The advantage of using a harmonic potential are two-fold. First, the confinement due to the harmonic potential can be easily tuned by varying a single parameter. Second, with respect to the other confinement options, the harmonic confinement is more similar to the confinement of electrons in experiments involving homogeneous magnetic fields \cite{Taut_PhysRevA_1993_v48_p3561,Taut_1994,Pino_2001,zeeman_split,ludena_hooke_h2}.
We focus here on quasi-one-dimensional systems with up to four electrons, which we will allow us to use the very accurate Complete Active Space Self-Consistent Field approach (CASSCF), to study the localization of the electrons.
We will study various properties to characterize the Wigner localization such as the one-body density and the particle-hole entropy.
The article is organised as follows. In section~\ref{Sec:Theory} we outline the theoretical and methodological aspects of our approach.
In section~\ref{Sec:Results} we will show and discuss the main results obtained with our formalism.
Finally, in section~\ref{Sec:Conclusions} we will draw our conclusions.
Notice that we will use Hartree atomic units throughout the article ($\hbar = m_e = e = 4\pi \epsilon_0$ = 1).

%In a first step, we have seen that singlet and triplet states for two electron systems have similar density profiles in the weak confinement regime ($k\to 0$) while for large confinement regime ($k \to \infty$) the density of each spin state converges to the free particle density. Then, for high spin states, for all number of electrons, we have computed the particle-hole entropy which is linked to correlation effects \cite{occ_corr_1,occ_corr_2,corr_entropy} and have been able to connect two lowest entropy phases: the Wigner molecule and the Fermi gas. These entropy curves depend on the number of electrons and the potential parameter $k$ and, therefore, on the density of the system. In all cases we have observed there is a certain value of $k$ for which the particle-hole entropy is maximal, hence static correlation effects play a big role.

\section{Theory and Method}
\label{Sec:Theory}

\subsection{Quantum mechanical model}

In this work, we study systems composed of few electrons that interact via a Coulomb potential and which are confined in a one-dimensional harmonic well along the $x$ axis. 
The Hamiltonian of the system can hence be expressed as  
\begin{equation}
    H = -\frac{1}{2} \sum_{i=1}^N \nabla_i^2 + \frac{k}{2} \sum_{i=1}^N x_i^2 + \sum_{i=1,j >i}^N \frac{1}{r_{ij}} \label{hamiltonian}
\end{equation}
where $N$ is the number of electrons, $k$ is the harmonic constant and $r_{ij}$ is the distance between electrons $i$ and $j$.
As is well known~\cite{whittaker_watson_1927,flugge_marshall_1952}, in the pure one-dimensional case the Coulomb operator is singular in the origin of coordinate system.
In order to sidestep this problem, we solve the Schr\"odinger equation by using three-dimensional basis functions distributed along one unique direction.
The final energy is then obtained by subtracting the transverse components of the kinetic energy for each electron \cite{brooke_diaz-marquez_evangelisti_leininger_loos_suaud_berger_2018}. 
In this work, we will study the behavior of such quasi-one-dimensional system, in particular, in the Wigner regime in the low-spin and high-spin states using the accurate CASSCF formalism which is suitable, for a small number of electrons, to treat equally well both these spin states.
\subsection{Distributed Gaussian orbitals}
In order to properly describe the wave function of the system corresponding to the Hamiltonian in Eq.~(\ref{hamiltonian}), we use an equidistant grid of normalized gaussian functions along a segment of the $x$ axis centred around $x=0$.
This choice ensures that even for small values of $k$ (weak confinement) the basis set is sufficiently flexible to yield an accurate wave function.
The gaussian functions are given by
\begin{equation}
    \phi_{\mu} (\textbf{r};\alpha,\textbf{R}_{\mu}) = \left(\frac{2 \alpha}{\pi} \right)^{3/4} \exp(-\alpha (\textbf{r}-\textbf{R}_{\mu})^2) \label{gaussian_basis}
\end{equation}
where $\textbf{R}_{\mu}=(x_{\mu},0,0)^T$ is the position of the center of gaussian $\mu$.
The limits of the segment are defined with respect to the classical turning points of the highest energy level which are given by $\pm x_0(N,k) = \pm \left( \frac{(2N+1)^2}{k} \right)^{1/4}$. 
We place a basis function in each of these turning points as well as in $x=0$, which corresponds to the position of the minimum of the potential well, and we add $2m$ equidistant gaussians between the classical turning points and $m$ gaussians on each side of these points. Therefore, there is a total number of $M=3+4m$ basis functions and the distance between adjacent basis functions is given by $\delta=\frac{2 x_0}{m+1}$. Since all basis functions have the same exponent $\alpha$, the overlap between neighbouring functions is given by $S(\alpha,\delta)=\exp(-\alpha \delta^2/2)$. One may also rewrite the overlap function in terms of a dimensionless parameter $\xi = \alpha \delta^2$ as $S(\xi) = \exp(-\xi/2)$. 
This dimensionless parameter characterizes the resolution of the basis set.
Here we choose $\xi=1.0$ since it has previously been demonstrated that this yields an accurate description of the wave function~ \cite{1D_Wigner_Alejandro,2_3_d_wigner_crystal,wigner_mol_ring_2019,wigner_2021_toulouse,brooke_diaz-marquez_evangelisti_leininger_loos_suaud_berger_2018}. 
Therefore, the exponent $\alpha$ depends on the harmonic constant, the number of electrons and the number of basis functions according to $\alpha = \frac{\xi (m+1)^2}{4 x_0^2(N,k)}$.
\subsection{One-body density and the particle-hole entropy}
In order to characterize the Wigner localization we will consider two properties of the system, namely the one-body density and the particle-hole entropy. The former is defined as
\begin{equation}
    \rho(\textbf{r}) = \sum_i \gamma_i \varphi_i^*(\textbf{r})\varphi_i(\textbf{r}) \label{density}
\end{equation}
where $\varphi_i(\textbf{r})$ are the natural orbitals and the coefficients $\gamma_i$ are the natural occupation numbers which can be linked to the amount of electron correlation in a system.~\cite{Giesbertz_2013,DiSabatino_2015}
Although the Wigner localization would emerge very clearly from an electronic density profile, it is cumbersome to use the density to study the variation of the localization as a function of the confinement parameter $k$.
Moreover, it is difficult to use the density to pinpoint the transition between the Fermi-gas and Wigner regimes.
For these reasons, we will also analyse the particle-hole entropy which is defined as 
\begin{equation}
    S = - \sum_i \gamma_i \log \gamma_i +  (1-\gamma_i) \log (1-\gamma_i) 
    \label{electron_hole_entropy}
\end{equation}
and whose behavior can be easily studied as a function of $k$.

% \textcolor{magenta}{\bf begin delete}
% At strong confinement (large $k$) the system is weakly correlated and can be described by a single Slater determinant.
% Therefore, the occupation numbers $\gamma_i$ will be either one or zero and the corresponding particle-hole entropy will vanish.
% Instead, at weak confinement (small $k$) the system is strongly correlated and, in general, the system has to be described by a multitude of Slater determinants. 
% However, in the high-spin state, thanks to the Pauli principle, we expect that a single Slater determinant would be sufficient to describe the system also in the small $k$ limit.
% Therefore, the particle-hole entropy would vanish in both the weak and strong confinement limits.
% Since the particle-hole entropy is non-negative, it would have to reach a maximum for some value of $k$.
% This value could then be interpreted as the transition point between the delocalized Fermi regime and the localized Wigner regime.
% \textcolor{magenta}{\bf end delete}

At strong confinement (large $k$) the system is weakly correlated and can be asymptotically (infinite $k$) described by a single Slater determinant.
Therefore, the occupation numbers $\gamma_i$ will be either one or zero, and the corresponding particle-hole entropy will vanish. This is true regardless the spin state of the system.
At weak confinement (small $k$), on the other hand, the system starts becoming correlated and, in general, has to be described by a multitude of Slater determinants. 
Therefore, the particle-hole entropy increases.
At very low values of k the non-dynamical correlation becomes important, i.e., several Slater determinants are equally important to describe the wave function.
In such a case, the structure of the wave function, and hence its entropy, will depend not only on the value of the total spin $S^2$, but also on its projection $S_z$.
However, in the high-spin state and for a maximum value of $|S_z|$, we expect that, thanks to the Pauli principle, a single Slater determinant will be sufficient to describe the system also in the small $k$ limit~\cite{Bloch}.
Therefore, the particle-hole entropy would vanish in both the weak and strong confinement limits.
Since the particle-hole entropy is non-negative, it would have to reach a maximum for some value of $k$.
This value could then be interpreted as the transition point between the delocalized Fermi regime and the localized Wigner regime.

Below, when discussing the results, we will show that this is indeed the case.
\section{Results and discussion}
\label{Sec:Results}
\subsection{Two electron system}
For a system of two electrons we set $m=25$ yielding a total of 103 gaussian basis functions centred around the bottom of the harmonic well. With this basis set, we have computed the singlet and triplet spin states of the system for several values of the confinement parameter $k$ and with different sizes of the active space. From these calculations we have concluded that the optimal active space is the one corresponding to CASSCF(2,5).
With this active space, we avoid convergence problems arising from small occupation numbers while still capturing the main electron correlation features.
We remind the reader that we use the standard notation for the active space, indicating by CASSCF(N,M) a calculation having an active space of $N$ electrons distributed within $M$ orbitals.

\begin{figure}[ht]
 \includegraphics[scale=0.45]{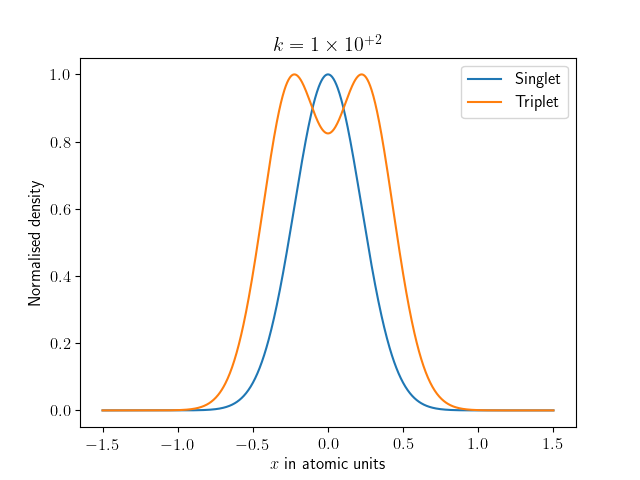}
  \includegraphics[scale=0.45]{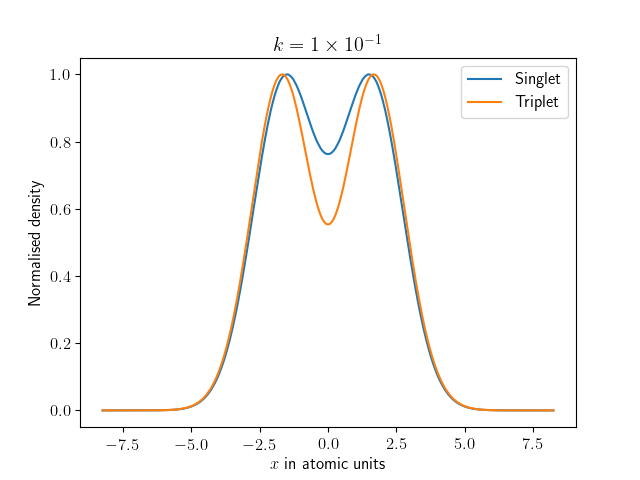}
  \includegraphics[scale=0.45]{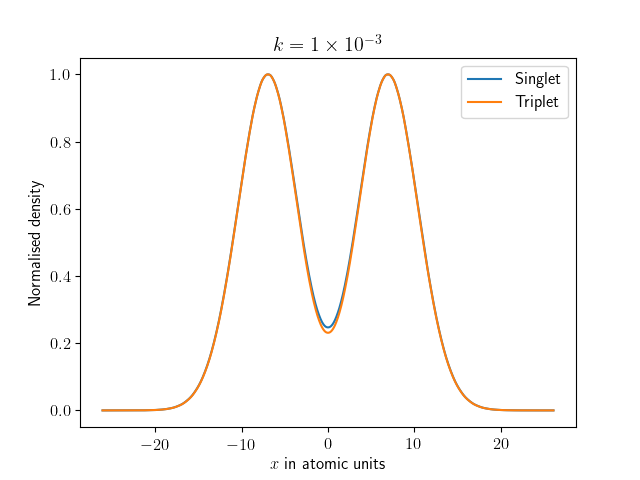}
  \includegraphics[scale=0.45]{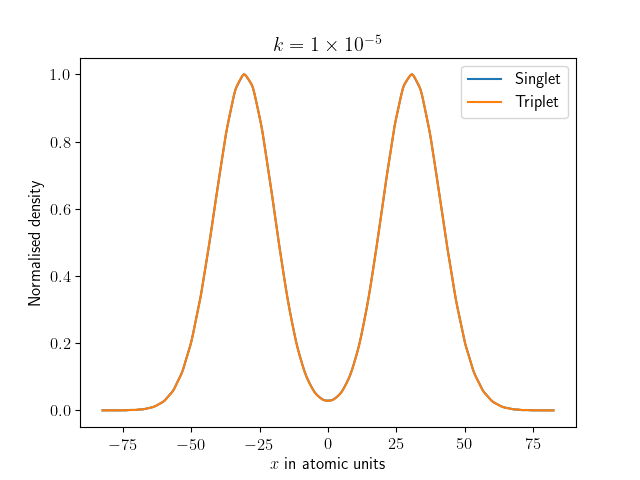}
    \caption{Normalised density profiles for singlet (blue) and triplet (orange) states for several confinement strengths $k$. 
    Notice in the weak confinement limit, both profiles are alike and they differ in the strong confinement limit.}
    \label{fig:densities_2_el}
\end{figure}

Once we have set the optimal level of theory, we have computed the singlet and triplet spin states for several values of the harmonic confinement parameter $k$ spanning several orders of magnitude. For each value of $k$ and for each spin state, we have analysed the resulting one-body density of the electrons.
We show some illustrative density profiles in figure \ref{fig:densities_2_el}.
As can be seen from this figure the density profile depends on both the confinement parameter $k$ and the spin state of the system.
As expected, for large values of $k$ the singlet state has one peak in its density profile while the triplet state has two peaks because of Pauli exclusion. Instead, for small values of $k$ both the singlet and triplet states have two peaks and the density profiles are very similar. For very large values of $k$ the density between the two peaks vanishes and we clearly see the localization of the electrons, this is the emergence of the Wigner regime.

\begin{figure}[ht]
    \centering
    \includegraphics[scale=0.5]{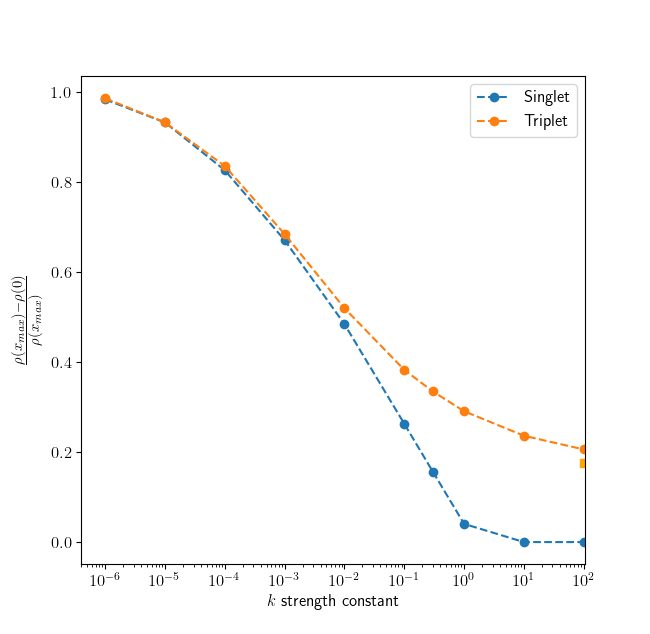}
    \caption{Variation of the location parameter with respect to the confinement parameter $k$ for singlet and triplet states, the orange square represents the asymptotic limit for location index of the triplet state.}
    \label{fig:height_var}
\end{figure}

By comparing the density evaluated at the maxima $\rho(x_{max})$ and the density evaluated at the minimum of the potential well $\rho(0)$, we may define a measure of localization as the difference of these two densities relative to the density evaluated at the maximum, i.e. $\frac{\rho(x_{max})-\rho(0)}{\rho(x_{max})}$. We will call this quantity the localization density which by definition takes values between zero (fully delocalized) and one (fully localized).
The values we obtained for this quantity are given in figure~\ref{fig:height_var} as a function of $k$. 
We observe that for small values of $k$ (smaller than $10^{-6}$) both spin states have similar localization densities and they are close to one, meaning that the electrons are fully localized and we can say that there is the formation of a Wigner molecule. As $k$ increases, the localization density of both spin states drops at the same rate up to at a certain point around $k=10^{-4}$ where they split. From this point on, the localization density decreases faster for the singlet state than for the triplet state and at a certain point $k\in(1,10)$ it drops to zero for the singlet state while the localization density for the triplet state converges asymptotically to a finite value given by $1-\frac{e^{1/2}}{2} \approx 0.1756$. This value corresponds to the localization density of two independent harmonic oscillators one in the ground state and the other in the first excited state.
This corresponds to the fact that that for large values of $k$ the one-body energy is much larger than the two-body energy and, therefore, the correlation effects in the system are less relevant.

\begin{figure}[ht]
    \centering
    \includegraphics[scale=0.45]{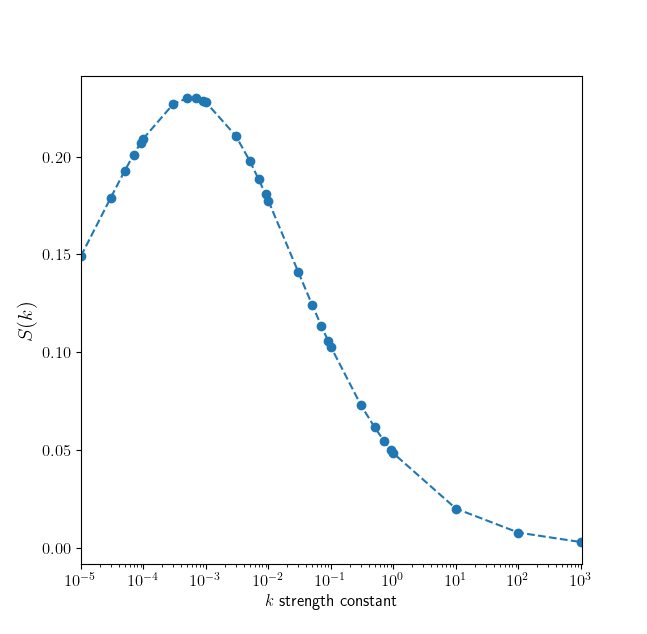}
    \caption{The particle-hole entropy for a two-electron system in the high-spin state as a function of the confinement parameter $k$. The maximum of the entropy is found for a value around $k=5\times10^{-4}$}
    \label{fig:entropy_2_el}
\end{figure}

With the aim of gaining a deeper understanding of the multi-determinantal character of our system, we have studied the particle-hole entropy as a function of the confinement parameter $k$ for the triplet state. The results we obtained are reported in figure \ref{fig:entropy_2_el}.
%As it can be seen, we have obtained points in $k$ for which the multi-determinantal nature starts to arise, i.e. around $k=1\times 10^{-5}$, reaches a maximum around $k=1\times 10^{-3}$ and then goes to zero for large values of $k$.
%
For large values of $k$ the particle-hole entropy tends to zero.
This is consistent with the fact that in the strong-confinement limit the independent particle model is a good description of the system and the resulting Fermi gas can be described by single Slater determinant.
We note that the shape of the curve we observe for large values of $k$ is coherent with results reported in previous works \cite{cioslowski:12jcp,cioslowski:14jcp}.
By decreasing the value of $k$ and, thereby reducing the confinement, the entropy increases which corresponds to an increase in the multi-determinantal character of the system.
We see that the entropy reaches a maximum around $k=5\times 10^{-4}$ and then decreases for smaller values of $k$.
% As we discussed before, for large $k$ values the independent particle model is a rather good description of our system and, based on the particle-hole entropy, the system can be represented by a single determinant. 
% Therefore, and the particle-hole entropy will be zero. O
This is consistent with the picture described before that in the limit $k \to 0$ the electrons localize, and since the system is in the high spin configuration, it can also be described with a single Slater determinant.
Therefore, in the $k\rightarrow 0$ limit, the particle-hole entropy of the Wigner molecule is expected to vanish.
Hence, we may infer the following behaviour: the two limiting cases i.e. the Wigner molecule ($k \to 0$) and the Fermi gas $(k \to \infty)$ are mono-determinantal states and hence their particle-hole entropy is zero.
Between these two extremes a maximum entropy is reached that corresponds to a state in which the multi-configurational character of the system is maximal. This point corresponding to $k=5\times 10^{-4}$ could be seen as a transition state that connects the Wigner molecule to the Fermi gas.
\subsubsection{A minimal-basis model}
In order to analyze the behavior of the two-electron system as we modify the confinement parameter, we have set up analytical models with a minimal number of gaussian functions for both the singlet and triplet states. 
For the singlet state we have used a single gaussian function centered at the minimum of the harmonic potential given by
\begin{equation}
    \sigma_0 (\textbf{r};\alpha_0) = \left(\frac{2 \alpha_0}{\pi} \right)^{3/4} \exp(-\alpha_0 \textbf{r}^2) \label{one_gauss}.
\end{equation}
in which $\alpha_0$ is the exponent to be optimized such that the lowest total energy is obtained.
This gaussian represents the doubly occupied orbital $^1\Sigma_g=\sigma_0^{\uparrow\downarrow}$.

Instead, for the triplet state we use two gaussian functions, each centred on one of the two classical equilibrium positions.
Since these basis functions are not orthogonal, we may adapt them by symmetry, by computing the even (gerade) and odd (ungerade) linear combinations, and then normalize them. 
As a result, we obtain an even function $\sigma_g$ and an odd one $\sigma_u$ given by, respectively,
\begin{align}
    \sigma_g (\textbf{r};\textbf{r}_0,\alpha_1)  \; & = \; \frac{1}{(2\pi)^{1/4} \sqrt{\frac{e^{2\alpha_1 x_0^2}+1}{\alpha_1^{1/2}}}} \Bigl(e^{-\alpha_1({\bf r}-{\bf r}_0)^2} \, + \, e^{-\alpha_1({\bf r}+{\bf r}_0)^2} \Bigr) \label{sigma_g},
    \\
    \sigma_u (\textbf{r};\textbf{r}_0,\alpha_1)\; & = \; \frac{1}{(2\pi)^{1/4} \sqrt{\frac{e^{2\alpha_1 x_0^2}-1}{\alpha_1^{1/2}}}} \Bigl(e^{-\alpha_1({\bf r}-{\bf r}_0)^2} \, - \, e^{-\alpha_1({\bf r}+{\bf r}_0)^2} \Bigr) \label{sigma_u},
\end{align}
in which $\alpha_1$ is the exponent to be optimized such that the lowest total energy is obtained.
By using these functions, we can occupy each orbital with one electron with parallel spins obtaining a triplet state $^3\Pi_u=\sigma_g^{\uparrow}\sigma_u^{\uparrow}$.
In summary, we model the Fermi gas as two electrons occupying a single gaussian orbital with a singlet spin state and the Wigner molecule as two electrons occupying two different symmetry-adapted orbitals with a triplet spin state.    

For the singlet state in the minimal basis $\sigma_0$, the energy is given by
\begin{equation}
     E_S(\alpha_0,k) \; = \; \alpha_0 \, + \, \frac{k}{4\alpha_0} \, + \, 2 \sqrt{\frac{\alpha_0}{\pi}} \label{singlet_minimal_energy}
\end{equation}
where the three terms on the right-hand side are the kinetic energy, the external potential energy and the two-electron repulsion energy. We note that, as mentioned before, we have subtracted the transverse component of the kinetic energy.

If we neglect the third term on the right-hand side corresponding to the Coulomb repulsion, 
the non-interacting singlet solution is given by $\alpha_0 = \sqrt{k}/2$, yielding a total energy of $ E_{0S}(k) \, = \, k^{1/2} \, + \, k^{1/4}(2/\pi)^{1/2}$, which shows that this approximation is valid for large values of $k$. Therefore, although the Coulomb repulsion is divergent for large values of $k$, the corresponding wave function will asymptotically converge to that of the non-interacting harmonic oscillator. 
We note that $E_{0S}(k)$ is an upper bound to the exact singlet energy $E_S(k)$.
The complete minimization including the Coulomb repulsion gives a fourth degree polynomial equation in terms of ${\alpha_0}^{-1/2}$, i.e. $-\frac{k}{4}({\alpha_0}^{-1/2})^4+\frac{2}{\sqrt{\pi}}({\alpha_0}^{-1/2})+1=0 $. Considering Descartes's rule of signs, one realizes that this equation has a single real positive root, thus the optimal value of the exponent $\alpha_{0}$ is unique and can be obtained by using the quartic formula. 
Hence, when the exponent $\alpha_0$ is optimized for all confinement strengths $k$, the singlet energy is just a function of $k$.

We continue our analysis with the triplet state. We consider first the non-interacting triplet, whose energy is obtained if we do not take into account the inter-electronic repulsion. The triplet energy is given by the sum of the energies of the two lowest levels of the harmonic oscillator, i.e., $\frac{1}{2} \sqrt{k}$ and $\frac{3}{2} \sqrt{k}$, yielding $  E_{0T}(k) \; = \; 2\sqrt{k}$.
As mentioned before, we use the symmetry adapted orthonormal minimal basis given in Eqs.~\eqref{sigma_g} and \eqref{sigma_u}. 
The main advantage of using these basis functions is that they have even and odd parity with respect to the origin. 
Hence, since both the kinetic and potential energy operators are even, all off-diagonal elements are zero and, therefore, the one-body Hamiltonian is diagonal. 
The corresponding matrix elements of the kinetic and potential energy operators are gives by, respectively,
\begin{align}
    T_{g,g}(\alpha_1;x_0(k)) \; &= \; \frac{\alpha_1}{2}  \frac{(1-4 \alpha_1 x_0^2+e^{2 \alpha_1 x_0^2})}{1+e^{2\alpha_1 x_0^2}} \hspace{3mm} 
\\
    T_{u,u}(\alpha_1;x_0(k)) \; &= \; \frac{\alpha_1}{2}  \frac{(-1+4 \alpha_1 x_0^2+e^{2 \alpha_1 x_0^2})}{1+e^{2\alpha_1 x_0^2}} \hspace{3mm} 
\\
    V_{g,g}(\alpha_1,k;x_0(k)) \; &= \; \frac{k}{8 \alpha_1} \Bigl( \frac{1+e^{2\alpha_1 x_0^2}(1+4\alpha_1 x_0^2)}{1+e^{2 \alpha_1 x_0^2}} \Bigr) \hspace{3mm} 
    \\
 V_{u,u}(\alpha_1,k;x_0(k)) \; &= \; \frac{k}{8 \alpha_1} \Bigl( \frac{-1+e^{2\alpha_1 x_0^2}(1+4\alpha_1 x_0^2)}{-1+e^{2 \alpha_1 x_0^2}} \Bigr) \hspace{3mm} 
\end{align}
We note that, for a non-zero value for $x_0$, in the limit of large values of $\alpha$, the kinetic energy converges to $\frac{1}{2}\alpha_1$, which is the kinetic energy of an electron trapped in a one-dimensional gaussian box. 
Furthermore, also for large values of $\alpha$, the potential energy converges to $\frac{1}{2}kx_0^2$ which is the classical potential energy for a particle at position $x=\pm x_0$.

Finally, the last interaction to take into account is the electron-electron repulsion. For the triplet spin state, there are two contributions:  the Coulomb repulsion  $(gg|uu)$ and the exchange interaction $(gu|gu)$ written in chemist's notation given by equations \eqref{coul_int}  and \eqref{exc_int} respectively. Evaluating the corresponding integrals, the interelectronic interaction is given by
\begin{eqnarray}
(gg|uu) &=& \iint \sigma_g^{*}(\textbf{r}_1) \sigma_g(\textbf{r}_1) \frac{1}{r_{12}} \sigma_u^{*}(\textbf{r}_2)\sigma_u(\textbf{r}_2) d \textbf{r}_1 d \textbf{r}_2  \label{coul_int} \\
(gu|gu) &=& \iint \sigma_g^{*}(\textbf{r}_1) \sigma_g(\textbf{r}_2) \frac{1}{r_{12}} \sigma_u^{*}(\textbf{r}_1)\sigma_u(\textbf{r}_2) d \textbf{r}_1 d \textbf{r}_2 \label{exc_int}
\end{eqnarray}

\begin{equation}
    V_{ee} (\alpha_1; x_0(k))= (gg|uu)-(gu|gu) = \frac{e^{4 \alpha_1 x_0^2}}{e^{4 \alpha_1 x_0^2}-1}  \left(\frac{\text{erf}(2x_0 \alpha_1^{1/2})}{2 x_0} - 2 \sqrt{\frac{\alpha_1}{\pi}}e^{-4 \alpha_1 x_0^2} \right) \label{ee_interaction}
\end{equation}
The first term in Eq.~\eqref{ee_interaction} corresponds to the Coulomb repulsion while the second one (which is rapidly decreasing with the distance between the centres) is the exchange interaction. Also for $V_{ee}$, one observes that for a given non-zero value of $x_0$, in the large $\alpha$ limit, the corresponding electron-electron interaction energy is given as $V_{ee}=\frac{1}{2x_0}$ which is the classical energy of two point particles located at positions $x=x_0$ and $x=-x_0$.
We can thus finally express the total energy of the triplet state in our model as
\begin{equation}
    E_T = T_{gg}+V_{gg}+T_{uu}+V_{uu}+ V_{ee} \label{triplet_e}.
\end{equation}
Once again we have to minimize $E_T$ to find the optimal $\alpha_0$ for each confinement strength $k$. 
As the energy as a function of $\alpha$ is rather complex, we have adopted a numerical procedure to minimize the triplet energy for all $k$ by varying $\alpha$.

\begin{figure}[ht]
    \centering
  \includegraphics[scale=0.48]{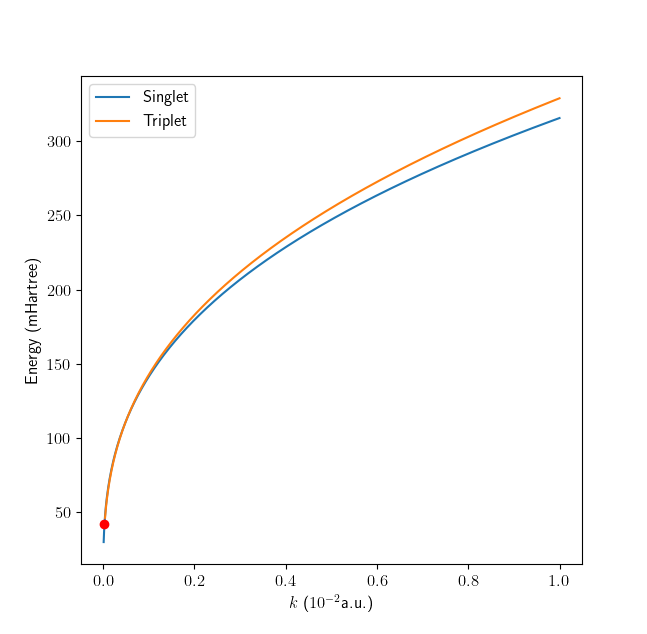}
        \includegraphics[scale=0.48]{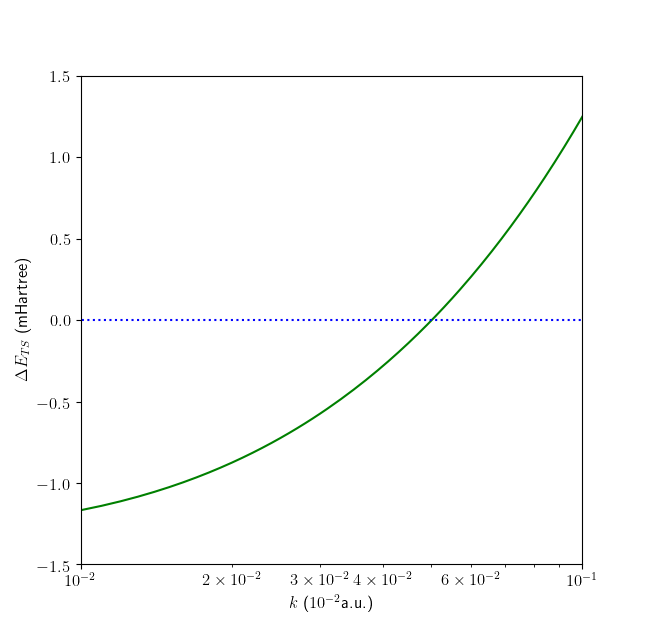}
        \caption{Left: energies for singlet and triplet spin states using minimal gaussian basis. Right: energy difference between the singlet and the triplet states for the two electron system using minimal basis. The zero energy gap happens around a value close to $k=5\times 10^{-4}$.}
    \label{fig:energies_k}
\end{figure}

In Fig.~\ref{fig:energies_k} we have reported the energies corresponding to the singlet and the triplet spin states as a function of the confinement strength $k$.
For small values of $k$, both states are very close in energy with the triplet state being slightly below the singlet state.
For some value around $k=5 \times 10^{-4}$ the energy curves cross and the triplet state is higher in energy for all $k$ beyond this point. 
It is convenient to recall that for large $k$, the energy of the triplet state goes as $E_{0T}=2k^{1/2}$ while the energy of the singlet state goes as $E_{0S}=k^{1/2}+(2/\pi)^{1/2}k^{1/4}$.

Thanks to the minimal basis model, we have shown that below a particular small value of $k$, the triplet state, which represents the  localized Wigner state, is lower in energy than the singlet state, which represents the delocalized state, and beyond $k\approx5\times 10^{-4}$ we find the opposite situation. 
We have thus been able to determine the order of magnitude of $k$ for which localization can occur.
Interestingly, the value of $k=5\times10^{-4}$) that we found for this transition using the model corresponds to the maximum of the particle-hole entropy.
\subsection{Systems of three and four electrons}
\begin{figure}[ht]
  \includegraphics[scale=0.48]{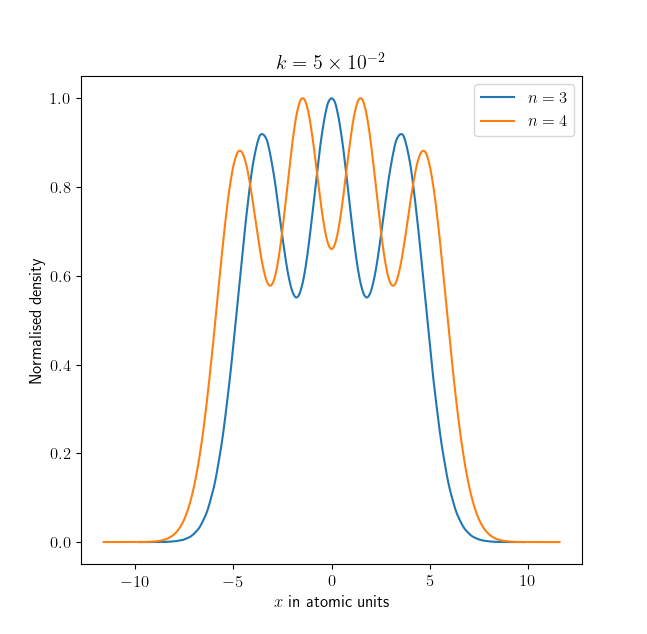}
  \includegraphics[scale=0.48]{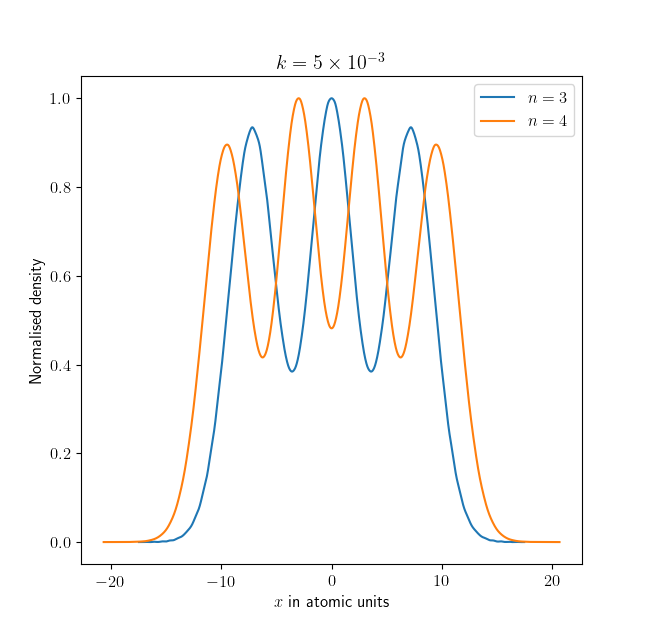}
  \includegraphics[scale=0.48]{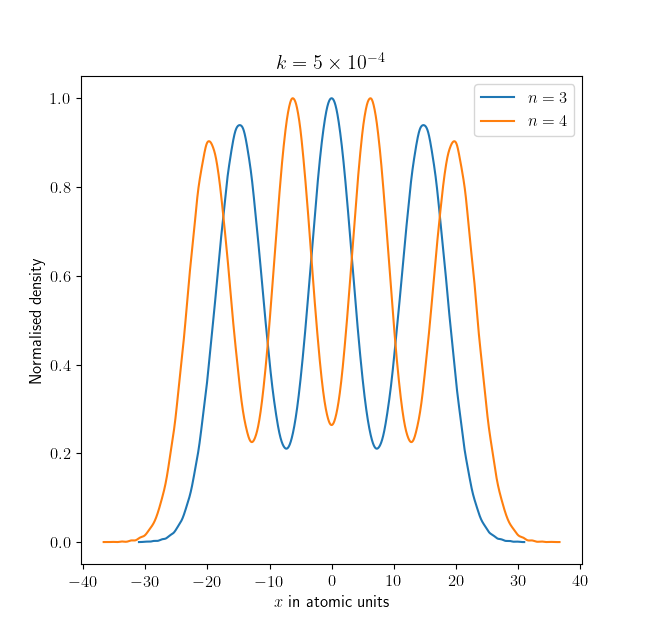}
  \includegraphics[scale=0.48]{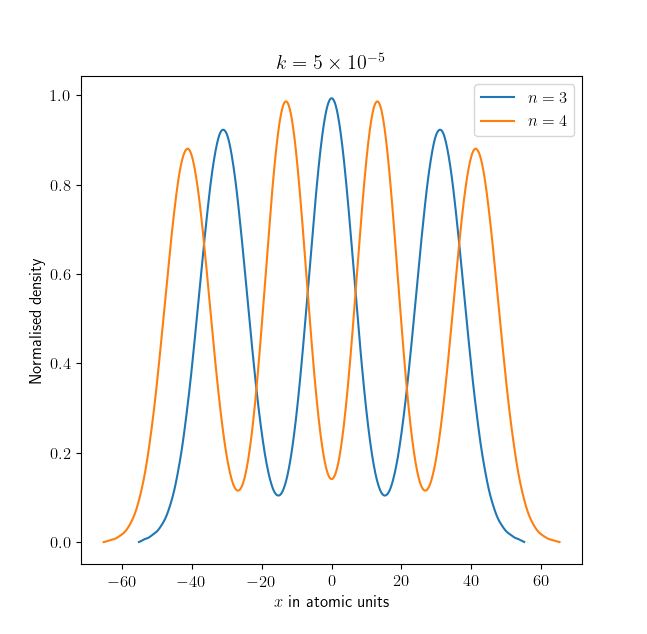}
    \caption{Normalised density profiles for three (blue) and four (orange) electron systems in high spin configuration for several confinement strengths $k$.}
    \label{fig:densities_3_4_el}
\end{figure}

\begin{figure}[ht]
    \centering
    \includegraphics[scale=0.7]{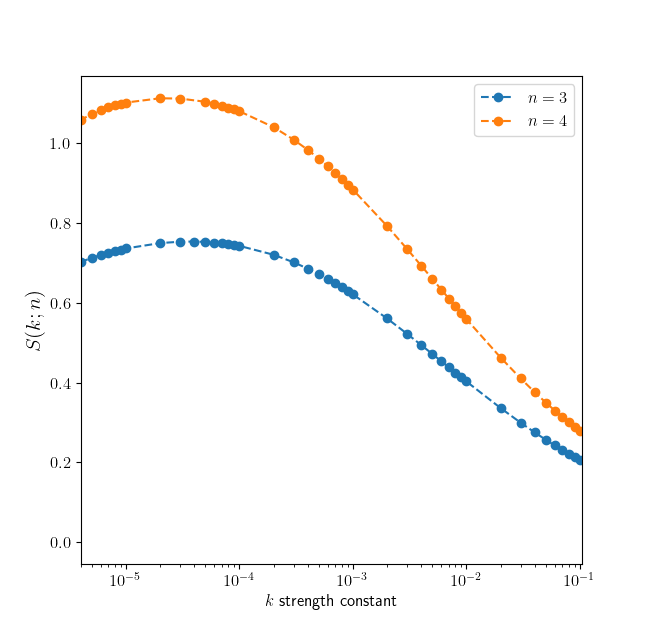}
    \caption{Computed particle-hole entropies for $n=3,4$ number of electrons with high spin state for several values of the confinement parameter $k$. }
    \label{fig:entropies}
\end{figure}

In this section we go beyond the two-electron case, in order to verify that the results obtained for systems with more electrons have similar localization properties as the two-electron system.
Since the high-spin state provided the clearest picture of the transition from the Fermi gas to the Wigner molecule we have constrained ourselves to this spin state in the calculation of three and four electrons.
We have performed CASSCF($N,2N$) calculations on these systems with the same grid of gaussian basis functions that we used for the two-electron case. 
We have computed both the density profiles for several values of $k$ as well as the particle-hole entropies as a function of $k$. The results we have obtained are reported in Figs.~\ref{fig:densities_3_4_el} and \ref{fig:entropies}.

We observe that the density profiles for three and four electrons are similar to that of the two-electron system.
By decreasing the value of the confinement parameter the peaks in the density profile become clearer and for very small values of $k$ the electrons localize.
We note that by increasing the number of electrons the localization occurs for slightly smaller values of $k$.
This was to be expected since the effective space that is available to the electrons is the same.
We obtain a similar picture when studying the particle-hole entropy.
Even though the curves for three and four electrons have a similar shape, the value of $k$ for which the entropy is largest depends on the number of electrons; the larger the number of electrons, the smaller the value of $k$ at which the entropy reaches its maximum. 
We infer that the localization properties of systems with $N>2$ electrons is similar to those of the two electron system and thus the conclusions are equivalent. We expect that similar results would be obtained for larger numbers of electrons.
We also conclude that we can define a transition state that connects the Wigner molecule and Fermi gas states and is characterized by $k_{max}(N)$ which corresponds to the maximal entropy point and it is a function of the number of electrons.
%The size and position of such transition state depends, of course, on the number of electrons.
%
\section{Conclusions}
\label{Sec:Conclusions}
In this work we have studied the localization of electrons in systems composed of two, three and four electrons confined to a one-dimensional harmonic potential characterised by a confinement parameter $k$ using accurate quantum-chemistry methods, i.e., CASSCF. 
To describe the many-body wave function we used basis sets composed of a large number of overlapping gaussian functions that were placed on a one-dimensional grid. 
To study the localization properties of the systems we calculated several properties such as the one-particle density and the particle-hole entropy.
From our study we conclude that the electrons localize to form a Wigner molecule when the value of $k$, and therefore the strength of the confinement, becomes sufficiently small.
Moreover, for certain values of $k$ a transition is taking place from the delocalized state to the localized state because the electron-hole entropy for the high-spin states shows a maximum for these values of $k$.
For two electrons this transition point occurs at $k=5 \times 10^{-4}$ while for systems with more electrons the transition point occurs for smaller values of $k$.

Finally, in order to analyze in more detail the results obtained for the two-electron system, we have derived analytical expressions using a minimal-basis set.
We have thus been able to model the Wigner-molecule regime as a triplet state with two identical gaussian basis functions and the Fermi gas regime as a singlet state with a single gaussian basis function. 
Using this scheme, we have minimized the energy of each system for several values of the confinement parameter $k$ and have seen that a crossing of the energies corresponding to the singlet and triplet wavefunctions occurs around $k=5 \times 10^{-4}$ which corresponds to the transition point found with the accurate CASSCF calculations.
This indicates that a change in the nature of the wavefunction must take place around this value and that the system enters the Wigner regime when $k<5 \times 10^{-4}$.

% Extending the study to high-spin states concerning three and four electrons systems, we have observed the same behaviour. In the two limit cases, the density and the particle-hole entropies correspond to the ones expected for the Wigner molecule $(k \to 0)$ and the Fermi gas ($k \to \infty$). By varying the confinement parameter $k$, we have computed a smooth path connecting these extreme states such that the particle-hole entropy reaches a maximum.
% For very small values of the confinement parameter, we have obtained some numerical instabilities and have not obtained reliable information on how the hole-particle entropy behaves in this regime. Further work concerning the study of this regime can be done in a close future. 

\section{Data Availability}

The data that support the findings of this study are available from the corresponding author upon reasonable request. 

\section{Acknowledgements}
As a first collaboration between the Laboratoire de Chimie et Physique Quantique from the University Paul Sabatier Toulouse III and Kimika Teoriko Taldea from the Univerity of the Basque Country, many organisations must be acknowledged.

On the one hand, this research was partly funded by Eusko Jaurlaritza (the Basque Government),
through Consolidated Group Project No. IT1254-19, PIBA19-0004, and 2019-CIEN-000092-01, and
the Spanish MINECO/FEDER 
Projects No. PGC2018-097529-B-100, PGC2018-098212-B-C21, EUIN2017-88605, and EUR2019-103825.

On the other hand, this work was partly supported by the French
``Centre National de la Recherche Scientifique'' (CNRS, also under the PICS action 4263).
It has received funding from the European Union's Horizon 2020 research and innovation program under the Marie Sk{\l}odowska-Curie grant agreement n\textsuperscript{o} 642294.
This work was also supported by the ``Programme Investissements d'Avenir'' under the
program ANR-11-IDEX-0002-02, reference ANR-10-LABX-0037-NEXT.

\providecommand{\latin}[1]{#1}
\makeatletter
\providecommand{\doi}
  {\begingroup\let\do\@makeother\dospecials
  \catcode`\{=1 \catcode`\}=2 \doi@aux}
\providecommand{\doi@aux}[1]{\endgroup\texttt{#1}}
\makeatother
\providecommand*\mcitethebibliography{\thebibliography}
\csname @ifundefined\endcsname{endmcitethebibliography}
  {\let\endmcitethebibliography\endthebibliography}{}

\end{document}